\documentclass[aps, prl, twocolumn, showpacs, floatfix,10pt,superscriptaddress]{revtex4-2}
\usepackage[utf8]{inputenc}
\usepackage{amsmath}
\usepackage{amsfonts}
\usepackage{amsbsy}
\usepackage{amssymb}
\usepackage{graphicx}
\usepackage{textcomp}
\usepackage[caption=false,singlelinecheck=false]{subfig}
\usepackage{xcolor}
\usepackage{mathrsfs}
\usepackage{mathtools}
\usepackage{bm}
\usepackage{gensymb}
\usepackage{braket,wasysym}
\usepackage{float}
\usepackage{tikz}
\usepackage[colorlinks,linkcolor=blue,citecolor=red,filecolor=magenta,urlcolor=red,breaklinks]{hyperref}
\usepackage[bottom]{footmisc}
\usepackage{verbatim}

\hypersetup{colorlinks=true, urlcolor=blue, citecolor=red, pdfborder={0 0 0}}
\usepackage{breakurl}
\usepackage{natbib}

\allowdisplaybreaks

\newcommand{\tcr}[1]{\textcolor{red}{#1}}

\normalfont
\def\be{\begin{equation}}
	\def\ee{\end{equation}}
\def\bea{\begin{eqnarray}}
	\def\eea{\end{eqnarray}}

\begin{document}
\title{
%Hyperuniformity-driven delocalization transitions with mobility edge in one-dimensional random systems
Delocalization Induced by Enhanced Hyperuniformity \\ in One-Dimensional Disordered Systems
}

\author{Junmo Jeon}
\email{junmojeon@sophia.ac.jp}
\affiliation{Physics Division, Sophia University, Chiyoda-ku, Tokyo 102-8554, Japan}
\author{Harukuni Ikeda}
\email{harukuni.ikeda@yukawa.kyoto-u.ac.jp}
\affiliation{Yukawa Institute for Theoretical Physics, Kyoto University, Kyoto 606-8502, Japan}
\author{Shiro Sakai}
\email{shirosakai@sophia.ac.jp}
\affiliation{Physics Division, Sophia University, Chiyoda-ku, Tokyo 102-8554, Japan}

\date{\today}
\begin{abstract}
In one dimension, any disorder is traditionally believed to localize all states. We show that this paradigm breaks down under hyperuniform disorder, which suppresses long-wavelength fluctuations and interpolates between random and periodic potentials. In tight-binding chains, strong hyperuniformity induces a sharp delocalization transition and the emergence of mobility edges. The transition is identified by the generalized fractal dimension and corroborated by the scaling of localization length and transmittance. Hyperuniform disorder thus provides a general mechanism for engineering mobility edges and controlling transport in low dimensions.
\end{abstract}
\maketitle

\textit{Introduction}
Disorder plays a central role in determining quantum transport in condensed matter systems~\cite{cohen2016fundamentals,baranovski2006charge}. A paradigmatic example is Anderson localization, where destructive interference induced by random scattering confines all electronic states in low dimensions, irrespective of disorder strength \cite{anderson1958absence,abrahams1979scaling}. This long-standing view has established localization as an unavoidable outcome of randomness in low dimensions. However, it is now understood that correlated disorder can fundamentally alter this picture: specific long-range correlations can modify the energy dependence of localization lengths and even induce divergence at certain energies, thereby creating mobility edges \cite{de1998delocalization,lugan2009one,piraud2012anderson,izrailev1995hamiltonian,izrailev1999localization,izrailev2012anomalous}. Moreover, in quasiperiodic systems such as Aubry–André model and its generalizations, which lack periodicity yet retain long-range order, not only mobility edges but also power-law–scaled critical eigenstates have been found \cite{aubry1980analyticity,zhang2022lyapunov,cookmeyer2020critical,biddle2009localization,hiramoto1989new}. These exotic electronic states enable unconventional forms of quantum transport that is neither ballistic nor insulating \cite{roche1997electronic,jeon2021topological,kellendonk2015mathematics}. Such theoretical predictions have been supported by wave-propagation experiments in correlated random media, including photonic crystals and cold-atom arrays~\cite{sanchez2007anderson,billy2008direct,phillips1991localization,biddle2010predicted,luschen2018single,vasco2018anderson}.

Recent studies of amorphous and aperiodic structures have shown that not all forms of randomness are equivalent: large-scale density fluctuations provide a key metric distinguishing hidden hierarchies among disordered systems~\cite{torquato2016hyperuniformity,torquato2018hyperuniform}. Of particular interest is the hyperuniform disorder, where long-wavelength fluctuations are suppressed while short-range irregularities persist.
This concept has attracted growing attention across diverse contexts, including photonic materials~\cite{torquato2018hyperuniform,florescu2009complete,man2013isotropic,froufe-perez2016role,milosevic2019hyperuniform,aubry2020experimental,takemori2025photonic} and condensed matter systems~\cite{xie2013hyperuniformity,zhang2016the,lethien2017enhanced,oguz2017hyperuniformity,zheng2020disordered,sakai2022hyperuniform,sakai2022quantum,crowley2019quantum,hori2024multifractality,koga2024hyperuniform,koga2025critical,wang2025hyperuniform}. How suppressing long-range fluctuations affects localization and transport, and whether this can evade universal localization in low dimensions, remains an open and actively studied question ~\cite{klatt2025transparency,shi2022many,karcher2024effect,sgrignuoli2022subdiffusive,vanoni2025effectivedelocalizationonedimensionalanderson}.

In this Letter, we employ hyperuniformization, a systematic method that transforms conventional disorder into hyperuniform random potentials approaching a periodic limit via momentum-space high-pass filtering, to study localization and transport through fractal dimension and transmittance across hyperuniformity classes. Contrary to the traditional view that any weak disorder localizes all states, sufficiently strong hyperuniformity
can produce delocalization transitions and mobility edges while preserving short-range randomness. These transitions do not require thermodynamically sustained long-range correlation, as in quasiperiodic systems. Instead, the suppression of large-scale potential fluctuations restores approximately uniform correlations over a finite range, thereby enabling mobility edges even in low dimensions. %\tcg{We further show that the delocalized states induced by strong hyperuniformity are localized in momentum space.} \HI{This sentence could be removed, as the data is not shown in the main text.}
Our results establish hyperuniform disorder as a practical route to engineer mobility edges and control quantum transport in low-dimensional systems.

\textit{Hyperuniformization} Let us consider a one-dimensional tight-binding model with a random potential. The Hamiltonian is given by
\begin{align}
    \label{H}
    &H=-t\sum_{i=1}^{L}(c_i^\dagger c_{i+1}+h.c.) +\sum_i^L V_i c_i^\dagger c_i.
\end{align}
Here, $c_i^\dagger \ (c_i)$ creates (annihilates) a spinless fermion at site $i$, and we impose periodic boundary conditions, 
$c_{L+1}=c_1$.
$L$ is the system size and 
%$t>0$
$t=1$
is uniform hopping strength. 
%We set $t=1$. 
The random potential, $V_i$, is uniformly distributed in $[-W,+W]$, where $W>0$ is the level of disorder. In the Anderson localization framework, the uncorrelated random potential leads to 
%the 
complete localization
for any small $W/t$ \cite{abrahams1979scaling}. 

However, recent works have established that one can further classify the random potentials in terms of their long-range fluctuations \cite{torquato2016hyperuniformity}. In particular, random systems 
%whose long-range fluctuations are negligible 
with suppressed long-range fluctuations
are called disordered hyperuniform \cite{sgrignuoli2022subdiffusive,torquato2016hyperuniformity,torquato2018hyperuniform,chen2023disordered,karcher2024effect,vanoni2025effectivedelocalizationonedimensionalanderson}. The hyperuniform potential is characterized by $\lim_{k\to0} S(k)=0$, where $S(k)=\frac{1}{L}\left\vert\sum_{j=1}^L V_j e^{ikj}\right\vert^2$ is the structure factor \cite{torquato2016hyperuniformity}. There are three different hyperuniform classes associated with the power-law scaling behavior of the structure factor. Specifically, when $S(k\to0)\sim k^\alpha$, the classes I, II and III are defined by $\alpha>1$, $\alpha=1$ and $0<\alpha<1$, respectively \cite{torquato2016hyperuniformity}. Note that $\alpha$ indicates the strength of hyperuniformity as a larger $\alpha$ suppresses long-range fluctuations more strongly.

\begin{figure}[h]
    \centering
    \includegraphics[width=0.5\textwidth]{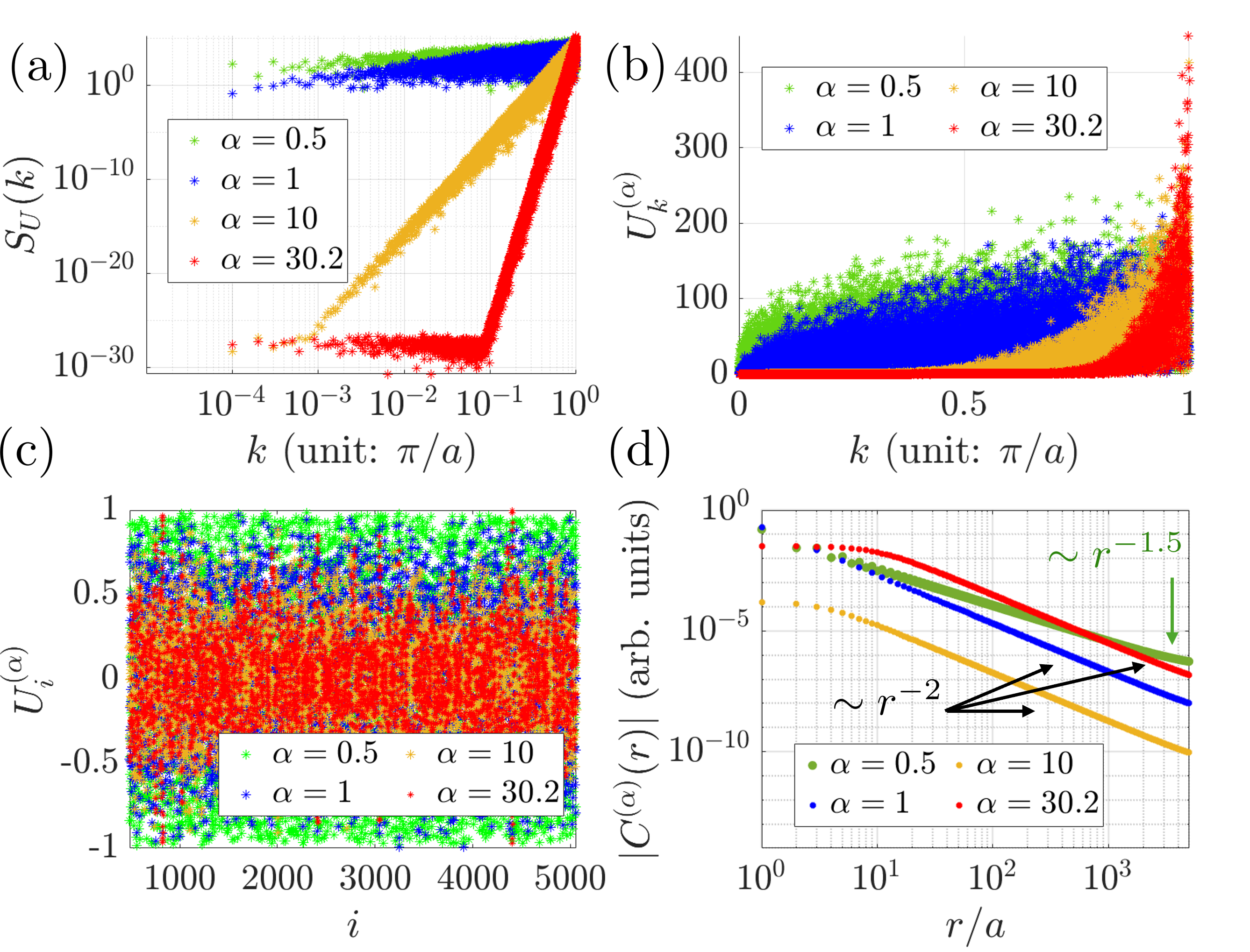}
    \caption{(a) Structure factor, $S_U(k)$ for $\alpha=0.5$ (class III), $\alpha=1$ (class II), $\alpha=10$ (class I), and $\alpha=30.2$ (class I).
    (b) Corresponding Fourier components of potentials, $U_k^{(\alpha)}$. 
    %for $\alpha=0.5,1,10$, and $30.2$.  
    %$a=1$ is the atomic distance. 
    (c) Hyperuniformized potentials, $U_i^{(\alpha)}$.
    %The green, blue, yellow, and red colors represent $\alpha=0.5$ (class III), $\alpha=1$ (class II), $\alpha=10$ (class I), and $\alpha=30.2$ (class I), respectively.
    (d) Two-point self-correlation function $C^{(\alpha)}(r)$ %(log-scale) 
    of 
    %hyperuniformized potentials for different $\alpha$ values.
    $U_i^{(\alpha)}$.
    For large $r$, $C^{(\alpha)}(r)\sim r^{-1.5}$ for $\alpha=0.5$, while $C^{(\alpha)}(r)\sim r^{-2}$ for $\alpha\ge 1$. %The black dashed and gray dotted lines indicate $r^{-1.5}$ and $r^{-2}$ scaling, respectively. 
    For $\alpha=30.2$, $C^{(\alpha)}(r)$ is approximately constant for $r/a\lesssim10$.
    %\HI{Adding a power-law fit for C(r) in Fig. 1(d) would make the behavior clearer, if possible.}
    %\HI{ $h_\alpha(k)$ is just a power-law function, and maybe not very informative. Can we plot $S(k)$ instead of $h_\alpha(k)$?}
    %\tcr{[**How about plotting (d) only for $k>0$? It is also preferable to use the same unit as (a) for the horizontal axis.**]}
    %\tcr{[**Please enlarge the fonts in figures.**]}
    }
    \label{fig: potentials}
\end{figure}

Based on the above definition, we systematically
generate hyperuniform potentials from a given random potential, $V_i$, keeping the level of disorder, $W$ \cite{karcher2024effect,klatt2025transparency,vanoni2025effectivedelocalizationonedimensionalanderson}. We refer to this procedure as hyperuniformization. Let us consider a momentum-space high-pass filter,
$h_\alpha(k)$, such that %$\lim_{k\to 0} h_\alpha(k)\propto \vert k\vert^{\alpha/2}$ for given $\alpha$.
$h_\alpha(k)\propto \vert k\vert^{\alpha/2}$ for small $k$.
As a concrete example, we define
\begin{align}
    \label{highpass}
    &h_\alpha(k)=\begin{cases}
        \vert k/k_c\vert^{\alpha/2}, & \vert k\vert\le k_c \\ 1, &  \vert k\vert >k_c.
    \end{cases}
\end{align}
Here, $k_c\le \pi/a$ is a constant, where $a=1$ denotes the atomic spacing. To explore the continuous evolution from a random %potential 
to a periodic system, we set $k_c=\pi/a$, so that $\lim_{\alpha\to\infty}h_\alpha(k)$ yields a periodic system \cite{supplementhyperuniformization1}. % It should be emphasized that our conclusions are independent of the particular functional form of the high-pass filter for $\vert k\vert\le k_c$ \tcr{[**We need to be careful with this statement since the conclusion may change between $k_c<\pi/a$ and $k_c=\pi/a$**]}.
We generate a hyperuniform potential, $U$, whose power of %scaling of the structure factor
the structure-factor scaling
is $\alpha$, as $U_i^{(\alpha)}=\mathcal{N}\mathcal{F}^{-1}(h_\alpha(k)\mathcal{F}(V_i))$, where $\mathcal{F}$ is Fourier transformation and $\mathcal{N}$ is the normalization constant to keep the maximum value of $|U_i^{(\alpha)}|$ at $W$.
Since the new potential $U$ has the structure factor $S_U(k\to0)=\frac{1}{L}\left\vert \mathcal{F}\left(U_i^{(\alpha)}\right)(k\to0)\right\vert^2\sim k^{\alpha}$, 
$U_i^{(\alpha)}$
is a hyperuniform potential whose class is determined by $\alpha$ [see Figs.~\ref{fig: potentials}(a) and \ref{fig: potentials}(b)]. Figure \ref{fig: potentials}(c) shows 
%hyperuniformized random
the potentials in real space. Symbols in green, blue, and yellow (red) denote class III, II, and I hyperuniform potentials, respectively.

Figure \ref{fig: potentials}(d) shows that the two-point correlation of potentials, $C^{(\alpha)}(r)=\vert \langle U_i^{(\alpha)}U_{i+r}^{(\alpha)}\rangle\vert$, where $\langle \ \rangle$ denotes the average over the site index $i$, diminishes for large $r$. %Specifically, $C^{(\alpha)}(r)=\frac{W^2}{3L}\sum_k\vert k\vert^\alpha e^{ikr}$.
%In the continuous limit, $C(r)\sim \frac{1}{\pi}\int_0^\pi k^\alpha \cos (kr)dk$. 
Specifically, for large $r$, $C^{(\alpha)}(r)\sim r^{-(\alpha+1)}$ for $\alpha\le 1$, while $C^{(\alpha)}(r)\sim 1/(\alpha^2+(\pi r/a)^2)$ for $\alpha>1$ \cite{supplementhyperuniformization1}.
%\HI{If this is correct $C^{(\alpha)}(0)$ should decrease on increasing $\alpha$. However, $C^{(\alpha=10)}(0)<C^{(\alpha=30.2)}(0)$ in Fig.~1(d). Please check.}
Note that $C^{(\alpha)}(r)$ is approximately constant for $r\lesssim a\alpha/\pi$, but %it 
decays in the power law as $C^{(\alpha)}(r)\sim r^{-2}$ for large $r$.
%For small $\alpha\le 1$, small $k$ components are dominant in the integral. Hence, extending the upper limit of the integral from $\pi$ to infinity does not introduce a significant error. This leads to $C(r)\sim r^{-(\alpha+1)}$. While, for large $\alpha>1$, $k^\alpha$ is concentrated on $k_c=\pi/a$. Thus, by using the substitution, $k=\pi-q$, where $q\ll\pi$, $C(r)\approx \frac{1}{\pi}\int_0^\pi (\pi-q)^\alpha \cos((\pi-q)r)dq$. Since $q\ll \pi$ for large $\alpha$, $(\pi-q)^\alpha\approx \pi^\alpha e^{-\alpha q/\pi}$. Hence, $C(r)\approx (-1)^r\pi^{\alpha-1}\int_0^\infty e^{-\alpha q/\pi} \cos(qr)dq$. Here, we have changed the upper limit of the integral based on the $e^{-\alpha q/\pi}$ factor. By applying the Laplace transform, we get $C(r)\sim \alpha/(\alpha^2+\pi^2r^2)$ for large $\alpha>1$. Note that $C(r)$ is approximately constant for $r<\alpha/\pi$, but it decays power-law as $C(r)\sim r^{-2}$ in the thermodynamic limit.
%For $\alpha\le 1$, $C(r)$ becomes a Dirac-$\delta$ function similar to that of the non-hyperuniform random potential \cite{anderson1958absence}.
%\tcr{[**Is it possible to exclude exponential with a large exponent?**]}
%While, for large $\alpha> 1$, $C(r)$ exhibits exponential decay. Specifically, $C(r)$ for $\alpha>1$ exponentially decays until $r\approx \alpha$, and for $r\gg \alpha$, it becomes negligibly small.
Thus, although there are differences in the presence or absence of correlations within a finite range depending on the hyperuniform class, the correlations of hyperuniformized potentials, created through Eq.~(\ref{highpass}) from a random potential, decay for large $r$ as far as $\alpha<\infty$, contrary to quasiperiodic systems, where the correlation is non-decaying \cite{supplementhyperuniformization1}. Hence, hyperuniformized potentials with finite $\alpha$ lack long-range order.
%\tcg{Thus, potential with any finite $\alpha$ exhibits correlations that decay algebraically. These correlations represent quasi-long-range order, distinct from the thermodynamically sustained long-range correlations of quasiperiodic systems.}
Nevertheless, we will show in the following
that variations in the degree of hyperuniformity with respect to $\alpha$ are capable of inducing a delocalization transition accompanied by the emergence of a mobility edge.
%, as we will show in the following.

%Before discussing the delocalization transition, let us point out the case where $\alpha$ approaches infinity. Since we are focusing on the case of $k_c=\pi/a$, $h_{\alpha}(k)$ converges pointwise to zero at all $k$ except $k=\pm\pi/a$, as $\alpha$ approaches infinity.
%\tcr{[**Do we need $m\ne 1$?**]}Hence, $\lim_{\alpha\to\infty}U_i^{(\alpha)}=(-1)^i W$, alternating periodic potential \cite{supplementhyperuniformization1}. Thus, our hyperuniformization procedure transforms a random potential into an alternating periodic one, pointwise in momentum space. This sets our approach apart from recent momentum-space filtering studies, which have considered stealthy ($k_c<\pi/a$) or weak (class II/III) hyperuniform disordered potentials, where delocalization remains constrained to finite-size systems \cite{klatt2025transparency,shi2022many,karcher2024effect,sgrignuoli2022subdiffusive,vanoni2025effectivedelocalizationonedimensionalanderson}.

\textit{Delocalization transition and mobility edge} Let us explore the localization characteristics of wavefunctions. To this end, we investigate the inverse participation ratio (IPR), $\mathcal{I}_n=\sum_i \vert \psi_n(i)\vert ^4$, where $\psi_n(i)$ is the normalized wavefunction of the $n$-th eigenstate at the site $i$ \cite{wegner1980inverse}. Generally, $\mathcal{I}_n\sim L^{-D_n}$, where $D_n$ is the fractal dimension \cite{fyodorov1993level}. In 
%a one-dimensional system,
one dimension,
$D_n=0$ (1) for localized (fully extended) states. When $0<D_n<1$, the state is called critical, neither fully extended nor exponentially localized \cite{bauer1990correlation}.

%\HI{Please explain briefly the reason for using DoIPR instead of IPR}

To study the localization characteristics of states in the vicinity of a given energy $E$, we consider the density of the IPR, $\mathcal{D}$ \cite{kramer1993localization,jeon2025electron}, at the given energy $E$ as
\begin{align}
    \label{DoIPR}
    &\mathcal{D}(E)=\frac{1}{\rho(E)\pi}\sum_n \mathcal{I}_n\frac{\Gamma}{(E-E_n)^2+\Gamma^2}.
\end{align}
Here, $\rho(E)=\pi^{-1}\sum_n \Gamma/[(E-E_n)^2+\Gamma^2]$ is the density of states and $\Gamma$ is half the linewidth. Note that $\mathcal{D}(E)$ is an averaged IPR of the states around a given energy $E$ by using the Lorentzian function. We use $\mathcal{D}(E)$ instead of IPR because in disordered systems with discrete spectra, individual IPR values fluctuate strongly.
We set $\Gamma=10^{-2}t$, which ensures sufficient averaging over discrete levels while preserving relevant spectral features.
From the scaling behavior $\mathcal{D}(E)\sim L^{-D_f(E)}$ for large system sizes, one can investigate $D_f(E)$, generalized fractal dimension at the energy $E$, given by
\begin{align}
    \label{FD}
    &D_f(E)=-\lim_{L\to\infty}\frac{\log \mathcal{D}(E)}{\log L}.
\end{align}
When the states around energy $E$ are ideally localized (extended), $D_f(E)=0$ (1).
While, $0<D_f(E)<1$ if either extended and localized states coexist or sufficiently many critical states appear near $E$. Thus, $D_f(E)$ quantifies
the average degree of delocalization of states around energy $E$.
%The generalized fractal dimension can be used to capture the presence of a mobility edge, since it generally depends on the energy \cite{jeon2025electron}.

\begin{figure}[h]
    \centering
    \includegraphics[width=0.5\textwidth]{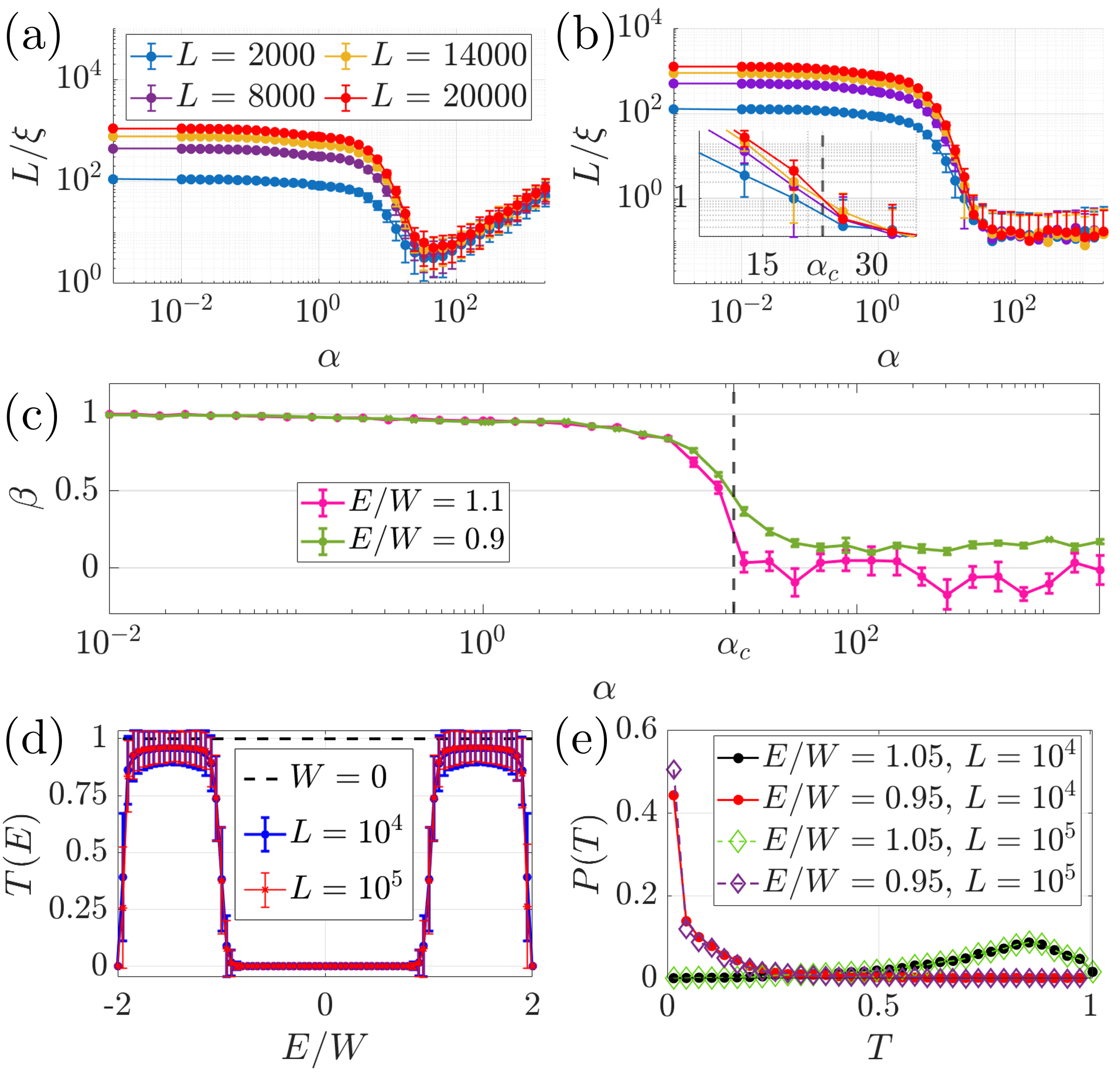}
    \caption{Scaling behavior of localization length, $\xi$ for (a) $E/W=0.9$ and (b) $E/W=1.1$. The black dashed line in the inset of (b) denotes
    the critical value $\alpha_c\approx 22$,
    where the scaling behavior falls into the ballistic regime, 
    i.e., 
    $\xi\sim L^1$ for $\alpha>\alpha_c$. (c) Scaling power, $\beta$ as a function of $\alpha$. (d) Transmittance of the system connected to two semi-infinite periodic leads as a function of $E/W$. Here, $\alpha=50$. The black dashed line represents the case of $W=0$. (e) Probability distribution of transmittance for different energies. We set $W=t=1$ and sampled $10^4$ different random potentials.  %\HI{Is $W=1$ for all panels (a)-(e)?}\tcc{Yes. W=1 is universal in this figure.}
    }
    \label{fig: transport}
\end{figure}

For small $\alpha$, we find that $D_f(E)$ remains small for all $E$, regardless of $W$. In the thermodynamic limit, $D_f(E)$ is expected to vanish, indicating localization of the states. This conclusion is supported by the localization length, which remains independent of system size in this regime, see Fig.~\ref{fig: transport}(c) \cite{supplementhyperuniformization1}.

%We first consider small $\alpha$. Figure~\ref{fig: fractal}(a) shows $D_f(E)$ at $\alpha=0.5$ for various disorder strength $W$. Note that $D_f(E)$ remains small for all $E$, which is essentially independent of $W$. In the thermodynamic limit, $D_f(E)$ is expected to vanish, indicating that the states are localized. This expectation is further supported by the localization length, which remains independent of system size in this regime, see Fig.~\ref{fig: transport}(c).

%We first focus on the case of small $\alpha$. Figure~\ref{fig: fractal}(a) shows $D_f(E)$ for different levels of disordere at small $\alpha$. We note that $D_f$ remains below 0.5 for all energies, independent of $W$ at small $\alpha\le1$. Thus, the states are either localized or at most weakly extended critical states for class II ($\alpha=1$) and class III ($0<\alpha<1$) hyperuniformity. Furthermore, it turns out that class II and class III hyperuniform potentials do not contribute to transport, as the localization lengths of the states are independent of system size, see Supplemental Material~\cite{supplementhyperuniformization1}. \HI{Which figure in SM?}

\begin{figure}[h]
    \centering
    \includegraphics[width=0.5\textwidth]{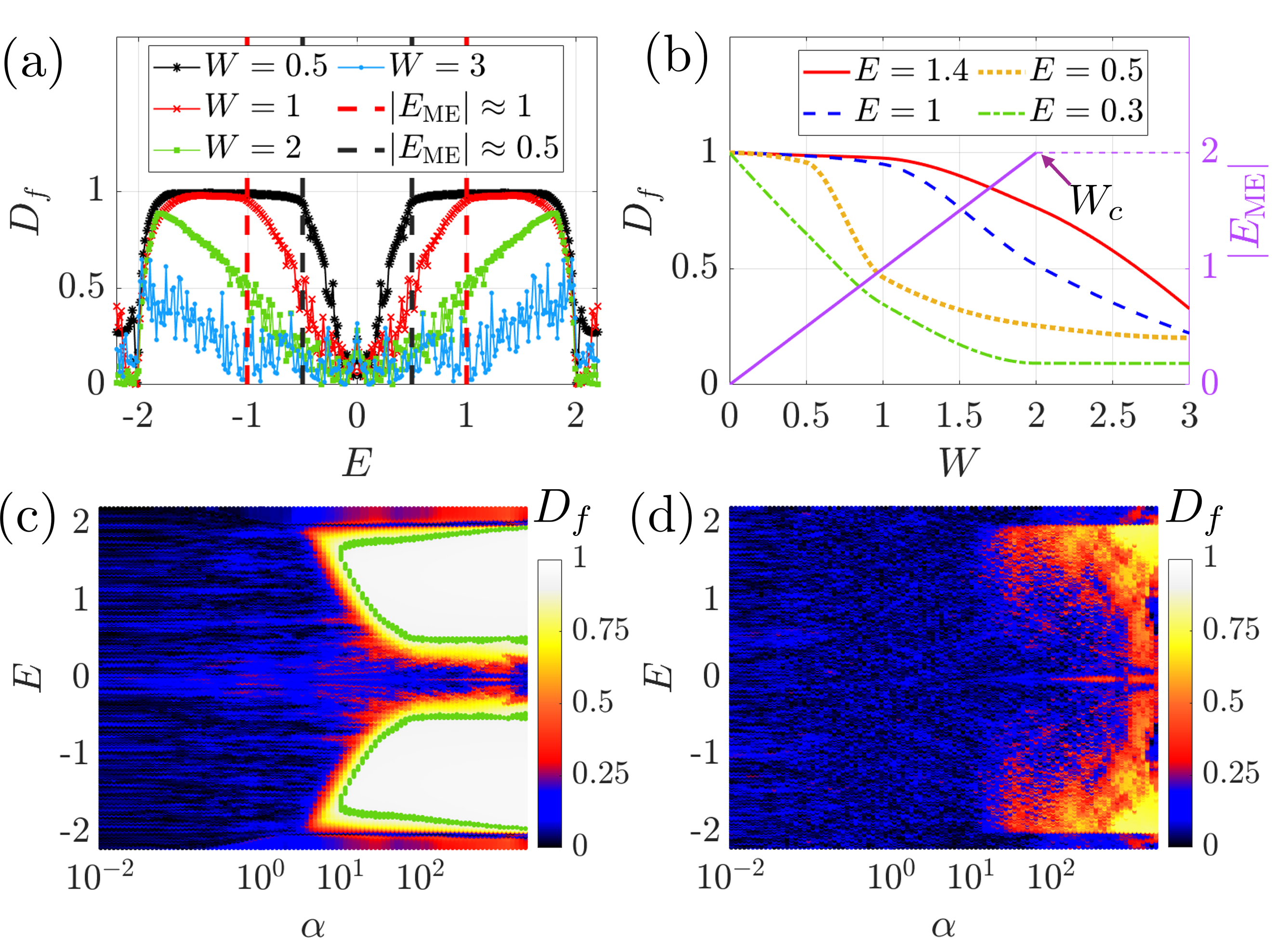}
    \caption{(a) $D_f(E)$ for different $W$ at $\alpha=100$. The black and red dashed lines are drawn to emphasize the mobility edges at $\vert E_{\mathrm{ME}}\vert=W$ for $W=0.5$ and $1$, respectively. (b) $D_f(E)$ plotted against $W$ at $\alpha=100$ for various $E$'s. The purple solid line (the right $y$-axis) denotes the mobility edge, $\vert E_{\mathrm{ME}}\vert=W<W_c=2$.
    Landscape of generalized fractal dimensions $D_f(E)$ plotted against the energy and strength of hyperuniformity, $\alpha$, for different levels of disorder: (c) $W=0.5$ and (d) $W=3$. The green curves in (c) denote the mobility edge. We use $10^3$ samples of random potentials and $2000\le L\le20000$. We set $t=1$.
    %(d) Schematic landscape of $D_f$ for sufficiently large $\alpha<L$. For $W/t<W_c/t=2$ (gray dashed line), the states for $W/t<\vert E\vert/t\le 2$ are extended, while the states for $\vert E\vert\le W$ are either critical or localized. The blue solid lines denote the mobility edges. For $W/t\ge2$, the states for $\vert E\vert/t\le 2$ are either critical or localized, with stronger localization for smaller $|E|$.
    %The gray dashed line is drawn to emphasize the critical level of disorder, $W_c/t=2$. 
    %The states for $\vert E\vert/t>2$ are either localized or 
    %weakly extended 
    %critical for all $W$.
    %\HI{The current panel (b) looks great, but its message is a bit unclear. Instead, how about showing $D_f$ for small and large $\alpha$ in panels (a) and (b), respectively? We could first explain, using panel (a), that the states are localized $D_f=0$ for small $\alpha$ across all $W$. Then, using panel (b) (corresponding to the current panel (a)), we can discuss that the mobility edge emerges for large $\alpha$.}
    }
    \label{fig: fractal}
\end{figure}

Now we consider large $\alpha$. Figure~\ref{fig: fractal}(a) shows $D_f(E)$ at $\alpha=100$ for various $W$. 
For weak disorder $W<2$, we find $D_f(E)\approx 1$ in the region $W< \vert E\vert\le 2$, indicating fully extended states~(black and red curves).  Whereas, for strong disorder $W\ge 2$, $D_f(E)<1$ for all $E$~(green and blue curves), indicating either localized or critical states. Notably, for $W<2$, $D_f(E)$ starts decreasing from unity near $E \approx W$ [Fig.~\ref{fig: fractal}(b)]. Thus, for sufficiently large $\alpha$, a mobility edge appears at $E_{\mathrm{ME}}=\pm2$ and $ E_{\mathrm{ME}}=\pm W$ [dashed lines in Fig.~\ref{fig: fractal}(a)], but only for $W<W_c=2$, see Fig.~\ref{fig: fractal}(b). The critical disorder $W_c=2$ and mobility edge $E_{\mathrm{ME}}=\pm 2$ originate from the bandwidth of uniform hopping $t=1$, while the mobility edge at $E_\mathrm{ME}=\pm W$ arises from the high-pass filter, which selects components near $k_c=\pi/a$ at large $\alpha$. Further details are presented in the Supplemental Material~\cite{supplementhyperuniformization1}.

Next, let us explore general $\alpha$ cases. Figures~\ref{fig: fractal}(c) and \ref{fig: fractal}(d) respectively show the landscapes of $D_f(E)$ for $W=0.5$ and $3$ as a function of $E$ and $\alpha$. For all $W$, the class-I hyperuniformity with large but finite $\alpha\gtrsim10$ leads to delocalization, where $D_f(E)$ begins to have a nonzero value except around $E=0$.
In contrast, for small $\alpha$, 
including class-II ($\alpha=1$) and class-III ($0<\alpha<1$) hyperuniformity, $D_f(E)\approx 0$ for all $E$ and $W$, indicating that the states are localized. Hence, only a strongly hyperuniform potential leads to delocalization. For $W<W_c$, large $\alpha\gtrsim10$ gives rise to the mobility edges at $W\le \vert E_{\mathrm{ME}}\vert<2$, which approach $W$ and $2$ as $\alpha$ increases~[Fig.~\ref{fig: fractal}(c)], whereas for $W>W_c$, the states are either localized or critical and the mobility edge does not appear~[Fig.~\ref{fig: fractal}(d)].

%This is because, in class II and III hyperuniform systems, as well as in class I systems with small $\alpha$, the randomness persisting at long range induces destructive interference of waves, thereby leading to localization of the same nature as Anderson localization \cite{karcher2024effect,vanoni2025effectivedelocalizationonedimensionalanderson}. In contrast, for class I hyperuniformity with large $\alpha$ that virtually shows stealthiness [see Fig.~\ref{fig: potentials}(a)] \cite{torquato2021structural}, most long-range fluctuations are suppressed, and the potential takes the form of rapidly oscillating within slowly varying envelopes---beatings.
%This allows delocalization to occur despite the effects of disorder.

\textit{Anomalous quantum transport} While the fractal dimension study shows hyperuniformity-induced delocalization and the emergence of mobility edges, analyzing the scaling of the localization length and the transmittance provides direct insight into the transport properties, linking delocalized states to experimentally measurable conduction features and providing a more complete picture of localization in hyperuniform disordered systems \cite{nazarov2009quantum,rammer2018quantum}.

%While the fractal dimension provides a clear indication of hyperuniformity-induced delocalization transitions and the emergence of mobility edges, a complementary understanding of transport properties is essential for characterizing the conduction behavior of the system. In particular, examining the scaling of the localization length and the transmittance offers direct insight into how the delocalized states contribute to experimentally measurable transport features, providing a more comprehensive picture of localization phenomena in hyperuniform disordered systems.

First, let us investigate the scaling behavior of the localization length, $\xi$, as a function of the system size $L$. To compute $\xi$, we use the transfer matrix method \cite{zhang1996method,mostafazadeh2020transfer}. Specifically, by rewriting the Schr\"odinger equation as \begin{align}
    \label{transfer}
    &\begin{pmatrix}
        \psi_{i+1} \\ \psi_{i}
    \end{pmatrix}=\begin{pmatrix}
        \frac{E-U_i^{(\alpha)}}{t} & -1 \\ 1 & 0
    \end{pmatrix}\begin{pmatrix}
        \psi_{i} \\ \psi_{i-1}
    \end{pmatrix}=T_i^{(\alpha)}(E)\begin{pmatrix}
        \psi_{i} \\ \psi_{i-1}
    \end{pmatrix},
\end{align}
we define the transfer matrix $M_L^{(\alpha)}(E)=\Pi_{i=1}^{L} T_i^{(\alpha)}(E)$. Then, the localization length is given by the inverse of the Lyapunov exponent, $\xi=1/\gamma$, where $\gamma=\lim_{L\to\infty} \log(\vert\vert M_L\vert\vert)/L$ \cite{barreira2017lyapunov,castanier1995lyapunov}. We first consider open boundary conditions,
$\psi_{0}=\psi_{L+1}=0$. The scaling behavior of $\xi$ indicates the insulating or ballistic regimes. In detail, $\xi\sim L^{1-\beta}$, where $\beta=1 (0)$ for insulating (ballistic) regime and $0<\beta<1$ for sub-ballistic transport~\cite{cohen2016fundamentals,abrahams1979scaling,kellendonk2015mathematics,roche1997electronic}.

When $\vert E\vert \le W$, we obtain $\xi<L$ for all $\alpha$ and $L/\xi$ increases as $L$ increases
[see Fig.~\ref{fig: transport}(a)]. For this energy window,  we have $0<\beta<1$ even for large $\alpha$~[see Fig.~\ref{fig: transport}(c)],
\textit{i.e.}, 
sub-ballistic transport with suppressed system-size dependence of $L/\xi$ ~\cite{supplementhyperuniformization1}. Whereas, for $W<\vert E\vert<2$ [Fig.~\ref{fig: transport}(b)], we have a critical $\alpha=\alpha_c$, where $L/\xi$ becomes equal to 1 [see black dashed line in the inset of Fig.~\ref{fig: transport}(b)]. The localization length becomes larger than the system size near this critical point, $\alpha_c\approx22$ for all $L$, and $L/\xi$ no longer exhibits a clear 
increase with $L$ for $\alpha>\alpha_c$, i.e., $\beta\approx 0$~[see Fig.~\ref{fig: transport}(c)]. This indicates that the system is metallic.
%Figure~\ref{fig: transport}(c) exhibits scaling power, $\beta$ as a function of $\alpha$ for different energies. For $\vert E\vert\le W$, $\beta$ remains positive, while $\beta\approx0$ for $W<\vert E\vert<2t$ at $\alpha>\alpha_c$.
%This indicates that the system is fully delocalized for $W<\vert E\vert<2t$. 
%Note that the sharp energy dependence of the localization length reveals that hyperuniformity induces a delocalization transition with a mobility edge at $E=\pm W$.

Next, we study the transport of a plane wave with energy $E$ through a disordered system coupled to a semi-infinite periodic chain. According to the Landauer formula \cite{imry1999conductance}, this directly corresponds to the conductance of the disordered quantum wire. Specifically, let us consider semi-infinite leads characterized by zero on-site potential and a uniform hopping amplitude, $t$, which is the same as that in the disordered region. The semi-infinite wires impose the constraint $\vert E\vert \le2t$. Given energy $E=2t\cos ka$, Landauer conductance \cite{kellendonk2015mathematics,jeon2021topological}
%, $G(E)$ 
is given by $G(E)=\frac{2e^2}{h}T(E)$
with
\begin{align}
    \label{T}
    &T(E)=\frac{4\sin^2 ka}{\vert M_{11}-M_{12}e^{ika}+M_{21}e^{ika}-M_{22}e^{2ika}\vert^2}.
\end{align}
Here, $M_L^{(\alpha)}=\begin{pmatrix} M_{11} & M_{12} \\ M_{21} & M_{22} \end{pmatrix}$ is the transfer matrix.
%        num = 4*sin(k)*sin(k);den = abs(M11-M12*exp(1i*k)+M21*exp(1i*k)-M22*exp(2*1i*k))^2;T = num / den;
%Note that for $W=0$, transmittance $T(E)=1-(E/2t)^2$, which goes to zero at $E=\pm 2t$ since the group velocity vanishes (dashed curve in Fig.~\ref{fig: transport}(c)). 
In class-I hyperuniform disordered systems with large $\alpha$, the delocalization transition leads to a clear mobility edge at $|E|=W$
that separates an insulating regime [$T(E)=0$] from a conducting regime [$T(E)>0$] [see Fig.~\ref{fig: transport}(d)]. Figure~\ref{fig: transport}(e) shows the difference in the distribution of transmittance across the mobility edge, as well as its size dependence. For $\vert E\vert<W$, $T(E)$ is concentrated on zero, while $T(E)$ has finite value for $W<\vert E\vert$. Furthermore, for $W<\vert E\vert$, the transmittance remains nearly independent of system size, manifesting the hallmark of one-dimensional ballistic transport \cite{abrahams1979scaling,roche1997electronic}. In contrast, the transmittance decreases as the system size increases 
%when
\tcr{for}
$\vert E\vert<W$. These are consistent with the fractal dimension results.

%Lastly, we emphasize that the 

\textit{Discussion} 
%\HI{Consider separating the discussion and conclusion sections, or moving some of the content to the Supplemental Material.}
We emphasize that the hyperuniformity-driven delocalization transition does not rely on long-range correlations~\cite{izrailev1995hamiltonian,izrailev1999localization,izrailev2012anomalous}. Note that $C^{(\alpha)}(r)$ decays more slowly for $\alpha < 1$ than for $\alpha \ge 1$ [see Fig.~\ref{fig: potentials}(d)] while the delocalization occurs only for the latter. In class-I hyperuniformity with large $\alpha$, long-wavelength fluctuations are strongly suppressed, and the potential becomes nearly uniform over a finite range $r\lesssim a\alpha/\pi$. This effective local periodicity promotes the formation of extended states, leading to delocalization for large $\alpha$ despite the presence of disorder and algebraically decaying correlations. By contrast, quasiperiodic models such as Aubry--Andr\'e represent the extreme case where the correlations do not decay with distance
~\cite{supplementhyperuniformization1,aubry1980analyticity}. Our results thus demonstrate that strong hyperuniformity, rather than the degree of long-range correlations, is the essential driver of delocalization. 

Notably, the delocalization transition in the current hyperuniformized potential does not signal the emergence of ergodic states as in the three-dimensional Anderson model, but instead marks a transition toward plane-wave states, strictly localized in momentum space. Across the transition, the level spacing ratio exhibits a clear reversal in scaling \cite{supplementhyperuniformization1}: for small $\alpha$, it decreases with system size, characteristic of localization, while beyond the transition, it sharply increases, reflecting the onset of level repulsion \cite{haldar2014level,liao2020many}. However, in striking contrast to the three-dimensional Anderson model—where ergodic states evolve toward Wigner–Dyson statistics \cite{argyrakis1992density}—the level spacing statistics here collapse into a $\delta$-function distribution at large $\alpha$ \cite{supplementhyperuniformization1}. Hence, the delocalized states with $D_f(E)\approx 1$ are in fact momentum-space localized plane waves. This agrees with our hyperuniformization, turning the potential into an alternating periodic one as $\alpha$ approaches infinity \cite{supplementhyperuniformization1}.

\textit{Conclusion} 
To conclude, hyperuniform disorder provides a general mechanism for delocalization transitions and the emergence of mobility edges even in one-dimensional systems, where conventional disorder inevitably localizes all states. By tuning the degree of hyperuniformity, we show that class-I hyperuniform potentials with large $\alpha$ lead to delocalization, as indicated by the fractal dimension and the scaling of localization length. Transport analysis reveals sharply defined mobility edges, separating insulating and metallic regimes, with the latter exhibiting ballistic transport that is independent of system size. Notably, these transitions arise from strong, effectively stealthy, hyperuniformity, which enforces nearly uniform correlations across a finite region, rather than relying on long-range correlations. It turns out that delocalization transitions in quasiperiodic models such as Aubry-André can be viewed as an extreme form of hyperuniformized potential.
Overall, hyperuniform disorder emerges as a general mechanism for delocalization in low-dimensional systems, providing a unifying framework for engineering mobility edges and controlling quantum transport in disordered materials.

In completing this work, we realized that the recent study by Vanoni et al. \cite{vanoni2025effectivedelocalizationonedimensionalanderson} addressed a related issue in the context of stealthy hyperuniform disorder. Studying the cases of the Heaviside-function high-pass filter with $k_c<\pi/a$, they obtained a large enhancement of the localization length with increasing the stealthiness of the potential, while they concluded that the system stays in a localized phase in the thermodynamic limit.
Although our potential with a large $\alpha$ virtually has the stealthiness, an important difference is in the choice of $k_c$.
By fixing $k_c=\pi/a$, our hyperuniformization converges pointwise to an alternating periodic chain, eliminating residual short-range fluctuations. As a consequence, we uncover a ballistic metallic phase in one dimension, which persists even for disorder strengths $W\sim\mathcal{O}(t)$, well beyond the weak-disorder regime considered in Ref.~\cite{vanoni2025effectivedelocalizationonedimensionalanderson}. This conclusion is consistently supported by our analyses of fractal dimension, transmittance distributions, and level-spacing statistics, which demonstrate that delocalization is not a finite-size artifact but a robust feature of the thermodynamic limit.

Our hyperuniformization procedure, a systematic construction that maps disorder onto the discrete Fourier spectrum, can be extended to general (quasi)periodic systems such as the Aubry–André model, offering a hyperuniformity perspective on delocalization transitions and mobility edges, with a promising avenue being the exploration of correlated physics in these hyperuniformized systems, including extensions to higher-dimensional lattices.

\section{Acknowledgement}
%\tcr{* Please fill in this section *}
We thank T. Ohtsuki for useful comments.
% KAKENHI (H.I.)
This work was supported by JSPS KAKENHI
Grant Numbers JP25H01397, JP25H10398 (SS), JP23K13031 and JP25H01401 (HI).

\bibliography{reference}

\end{document}

\title{Supplemental Material 
%of
\\
for
`Delocalization Induced by Enhanced Hyperuniformity \\in One-Dimensional Disordered Systems'}

\author{Junmo Jeon}
\email{junmojeon@sophia.ac.jp}
\affiliation{Physics Division, Sophia University, Chiyoda-ku, Tokyo 102-8554, Japan}
\author{Harukuni Ikeda}
\email{harukuni.ikeda@yukawa.kyoto-u.ac.jp}
\affiliation{Yukawa Institute for Theoretical Physics, Kyoto University, Kyoto 606-8502, Japan}
\author{Shiro Sakai}
\email{shirosakai@sophia.ac.jp}
\affiliation{Physics Division, Sophia University, Chiyoda-ku, Tokyo 102-8554, Japan}

\maketitle

\section{Two-point correlation function}
\label{sec:-1}
In this Supplemental Material, we analytically derive the scaling behavior of the two-point correlation function, $C^{(\alpha)}(r)=\vert \langle U_i^{(\alpha)}U_{i+r}^{(\alpha)}\rangle\vert$, where $\langle \ \rangle$ denotes the average over the site index $i$, of the hyperuniformized potentials derived from the momentum space high-pass filter given by Eq. (2) in the main text. Let us assume that $V_i$ is a white-noise random potential, whose variance is given by $\sigma^2=W^2/3$. Since $U_i^{(\alpha)}=\mathcal{N}\mathcal{F}^{-1}(h_\alpha(k)\mathcal{F}(V_i))$, where $\mathcal{F}$ is the Fourier transformation and $\mathcal{N}$ is the normalization constant to keep the level of disorder, we have $C^{(\alpha)}(r)=\frac{\sigma^2}{La}\sum_k\vert k/k_c\vert^\alpha e^{ikr}$ for $U_i^{(\alpha)}$. Without loss of generality, we let $a=1$. By using the symmetry, we have $C^{(\alpha)}(r)=\frac{\sigma^2}{\pi^{\alpha+1}}\int_0^{\pi/a} k^\alpha \cos (kr)dk$ in the continuous limit. 

For $\alpha\le 1$, small $k$ components are dominant in the integral since $\cos(kr)$ gives a rapid oscillation for large $kr$. Hence, extending the upper limit of the integral to infinity does not introduce a significant error. Thus, $C^{(\alpha)}(r)\approx \frac{\sigma^2}{\pi^{\alpha+1}}\int_0^\infty k^\alpha \cos (kr)dk$. This integral is given by $\int_0^\infty k^\alpha \cos (kr)dk=-\Gamma(\alpha+1)\frac{\sin (\pi\alpha/2)}{r^{\alpha+1}}$, where $\Gamma(x)$ is the Gamma function. Therefore, we have large-$r$ behavior of two-point correlation function as $C^{(\alpha)}(r)\sim r^{-(\alpha+1)}$ for $0<\alpha\le 1$.

On the other hand, for $\alpha>1$, $k^\alpha$ is concentrated on $k_c$. Thus, by using the substitution, $k=\pi-q$, where $q\ll\pi$, $C^{(\alpha)}(r)\approx \frac{\sigma^2}{\pi}\int_0^\pi (\pi-q)^\alpha \cos((\pi-q)r)dq$. Since $q\ll \pi$ for large $\alpha$, we can expand $(\pi-q)^\alpha=\pi^\alpha(1-q/\pi)^\alpha=\pi^\alpha e^{\alpha\ln(1-q/\pi)}$. By using $\ln(1+x)\approx x$ for $x\ll1$, we have $(\pi-q)^\alpha\approx\pi^\alpha e^{-\alpha q/\pi}$. Hence, $C^{(\alpha)}(r)\approx \frac{\sigma^2(-1)^r}{\pi}\int_0^\infty e^{-\alpha q/\pi} \cos(qr)dq$, where $(-1)^r$ originates from $\cos((\pi-q) r)=(-1)^r\cos(qr)$. Here, we have changed the upper limit of the integral based on the $e^{-\alpha q/\pi}$ factor. By applying the Laplace transform, we get $C^{(\alpha)}(r)\approx \sigma^2(-1)^r\alpha/(\alpha^2+(\pi r/a)^2)$ for $\alpha>1$. Here, we recover $a$. Note that $C^{(\alpha)}(r)$ is approximately constant for $r<a\alpha/\pi$, but it decays power-law as $C^{(\alpha)}(r)\sim r^{-2}$ for large $r$ in the thermodynamic limit for $\alpha<\infty$.

In short, we have a large-$r$ behavior of the two-point correlation function as
\begin{align}
    \label{twopointsupple}
    &C^{(\alpha)}(r)\approx\begin{cases} -\sigma^2\pi^{-\alpha-1}\Gamma(\alpha+1)\sin(\pi\alpha/2)(r/a)^{-(\alpha+1)}, \ \ &\mathrm{for \ 0<\alpha\le 1} \\ \sigma^2(-1)^r\frac{\alpha}{\alpha^2+(\pi r/a)^2}, \ \ &\mathrm{for \ \alpha>1}\end{cases}
\end{align}

Before we end this section, let us consider the case of the Aubry-André model, which is a well-known quasiperiodic potential\cite{aubry1980analyticity}. The Aubry-André potential is given by
\begin{align}
    \label{supple:AA}
    &V_i=\lambda \cos(2\pi\beta i+\phi),
\end{align}
where $\beta$ is an irrational number and $\lambda$ is a strength of potential. $\phi$ is a constant that we can set $\phi=0$ without loss of generality. It is known that $\lambda/t$ determines the localization characteristics of the states. Particularly, $\lambda/t=2$ is the delocalization transition point. From the trigonometry identity,
\begin{align}
    \label{cforAA}
    & V_iV_{i+r}=\lambda^2\cos(2\pi\beta i)\cos(2\pi\beta i+2\pi\beta r)=\frac{\lambda^2}{2}(\cos(2\pi\beta r)+\cos(\pi\beta(4i+r)).
\end{align}
Since the last term vanishes after we take the average over the site index $i$, we get
\begin{align}
    \label{cforAA_2}
    & C(r)=\frac{\lambda^2}{2}\cos(2\pi\beta r).
\end{align}
Thus, the quasiperiodic potential admits the oscillating but non-decaying long-range correlation. This is different from the case of hyperuniformized disordered potentials whose two-point correlations are power-law decaying as Eq.\eqref{twopointsupple}.

%Thus, although there are differences in the presence or absence of correlations within a finite range depending on the hyperuniform class, the hyperuniformized potentials, created through Eq. (2) in the main text from a random potential, do not have long-range self-correlations in the thermodynamic limit, regardless of $\alpha<\infty$, contrary to quasiperiodic systems, which have long-range correlated potentials.

\section{hyperuniformized
potentials for large $\alpha$ limit}
    \label{sec:0}
\begin{figure}[h]
    \centering
    \includegraphics[width=0.95\textwidth]{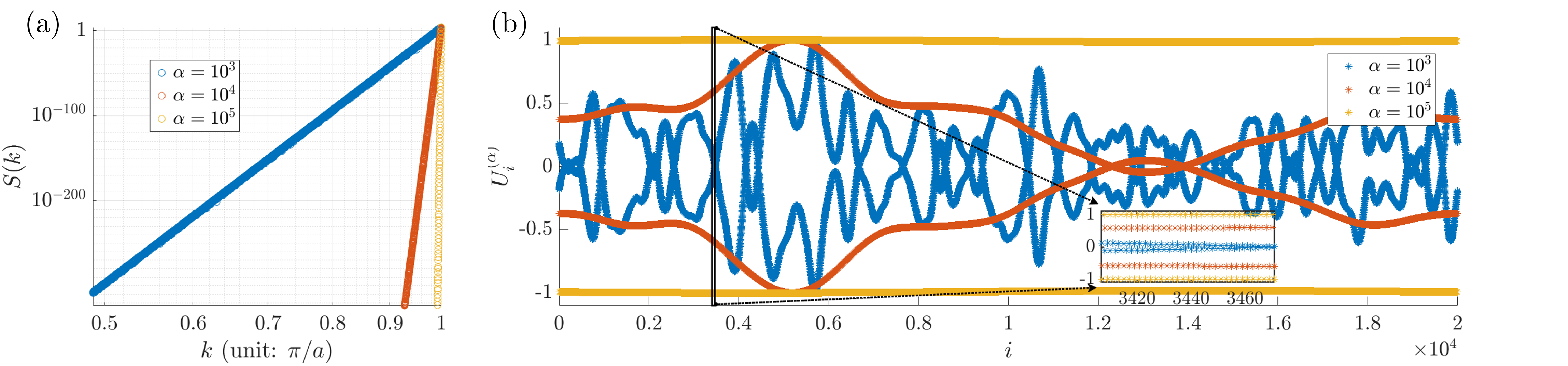}
    \caption{Hyperunfiromized potential for the case of $\alpha\gg1$. (a) Structure factor of hyperuniformized potential for $\alpha\gg1$. The blue, red, and yellow circles represent the cases $\alpha=10^3,10^4$ and $10^5$, respectively. (b) Hyperuniformized potential. The blue, red, and yellow stars represent the cases $\alpha=10^3,10^4$ and $10^5$, respectively. The inset is drawn to emphasize that the hyperuniformized potential for $\alpha=10^5$ becomes an alternating periodic chain. The system size, $L=20000$ and $W=t=1$.}
    \label{fig: Torquatosupple}
\end{figure}

Here, we discuss the case of $\alpha\gg1$. We note that $\lim_{\alpha\to\infty}h_\alpha(k)=0$ for $\vert k\vert<k_c$, while $\lim_{\alpha\to\infty}h_\alpha(k)=1$ for $\vert k\vert\ge k_c$. Particularly, $\lim_{\alpha\to\infty}h_\alpha(k_c)=1$. Since we set $k_c=\pi/a$, $h_\alpha(k)=0$ except $k=\pm\pi/a$. Thus, $S_U(k)=\frac{1}{L}\vert h_\alpha(k)\mathcal{F}(V_i)(k)\vert^2$ approaches zero except $k=\pm\pi/a$ as $\alpha$ approaches infinity [see Fig.~\ref{fig: Torquatosupple}(a)]. This is equivalent to halving the size of the Brillouin zone. Specifically, $\lim_{\alpha\to\infty} U_i^{(\alpha)}=(-1)^iW$ [see Fig.~\ref{fig: Torquatosupple}(b)]. Hence, the hyperuniformization procedure generates hyperuniform disordered potentials stemming from a random potential that are pointwise convergent to the alternating periodic potential. Note that for a finite-size system, $\alpha\gtrsim L$ is indistinguishable from a periodic system. This sets our approach apart from recent momentum-space filtering studies, which have considered stealthy ($k_c<\pi/a$) or weak (class II/III) hyperuniform disordered potentials, where delocalization remains constrained to finite-size systems \cite{klatt2025transparency,shi2022many,karcher2024effect,sgrignuoli2022subdiffusive,vanoni2025effectivedelocalizationonedimensionalanderson}.

\begin{figure}[h]
    \centering
    \includegraphics[width=0.95\textwidth]{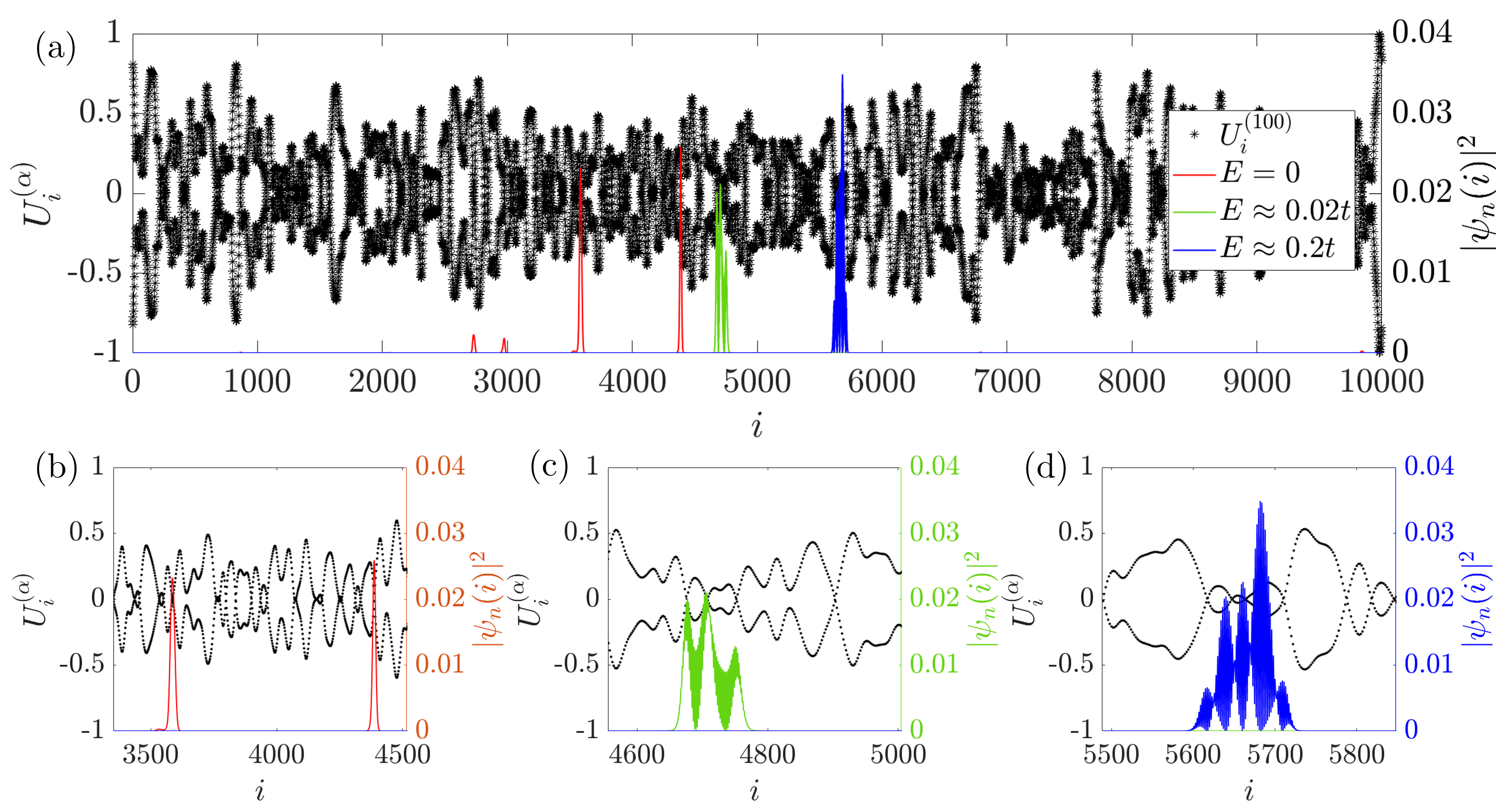}
    \caption{Localization for $\vert E\vert<W$ at large $\alpha$. (a) The red, green, and blue curves represent the localized states with $E=0$, $E\approx0.02t$, and $E\approx 0.2t$. The black symbol denotes the hyperuniformized potential for $\alpha=100$. (b-d) The zoomed-in plots for each localized state, along with the potential. The locations of the localized states coincide with the zeros of the envelopes of potential. The system size, $L=10000$ and $W=t=1$.}
    \label{fig: supplebeat}
\end{figure}

We note that the hyperuniformized potential, $U_i^{(\alpha)}$, exhibits rapid oscillations within slowly varying envelopes for sufficiently large $\alpha\gg1$ [see blue curves in Fig~\ref{fig: Torquatosupple}(b)]. This is because the high-pass filter, $h_\alpha(k)$ with sufficiently large $\alpha$, retains only the components near $k_c=\pi/a$. Figure~\ref{fig: supplebeat} shows that, even at large $\alpha$ and for weak disorder ($W<W_c$), localization for $\vert E\vert< W$ persists around the zeros of the slowly varying envelopes. In these regions, the states are confined by effective potential barriers whose height is at most $W$. Hence, the state whose energy $\vert E\vert\le W$ cannot delocalize into fully extended states, even under weak disorder ($W<W_c$) and strong hyperuniformity ($\alpha\gg 1$), due to the obstruction from these effective potential barriers.

\section{Spectrum of hyperuniformized disordered systems}
\label{spectrumsection}
\begin{figure}[h]
    \centering
    \includegraphics[width=0.95\textwidth]{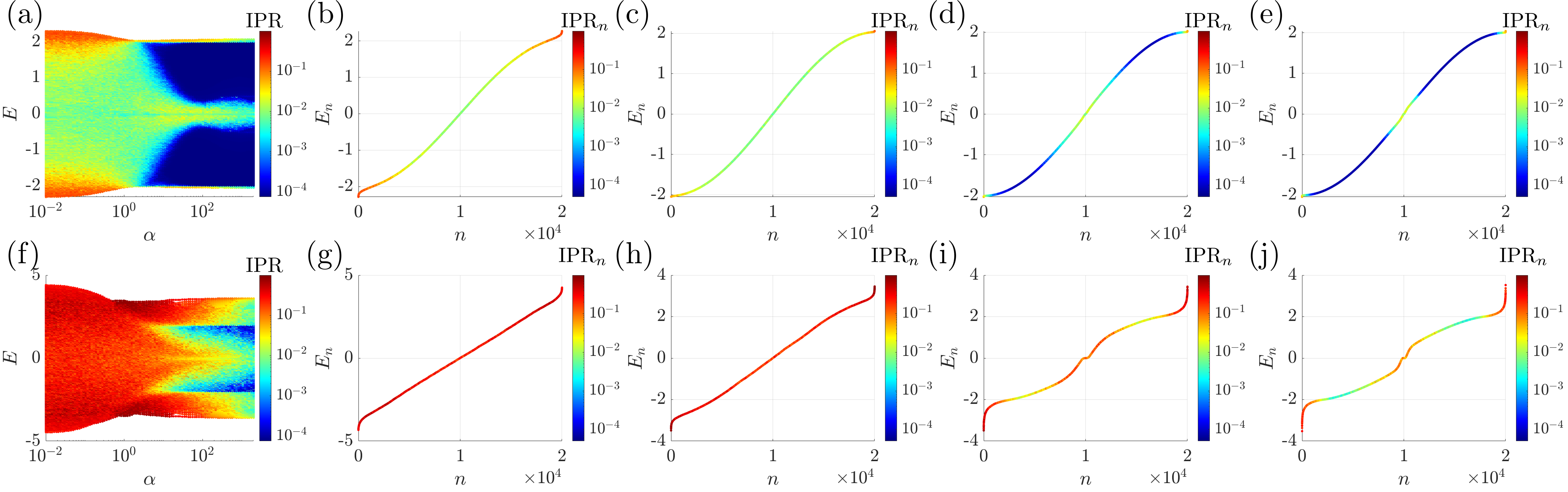}
    \caption{Landscape of the IPR and energy spectrum for (a-e) $W/t=0.5$ and (f-j) $W/t=3$. (a,f) Landscape of the IPR. The spectrum of various $\alpha$ values: (b,g) $\alpha=0.5$, (c,h) $\alpha=1$, (d,i) $\alpha=10$, and (e,j) $\alpha=30.2$, respectively. The colors of (b-e) and (g-j) represent the IPR values of each eigenstate. $L=20000$ and $t=1$.}
    \label{fig: supplespectrum}
\end{figure}

Here, we present the energy spectrum and the landscape of the inverse participation ratio (IPR) as a function of $\alpha$. Note that the IPR quantifies the strength of localization of the wavefunction. Figures~\ref{fig: supplespectrum}(a,f) show the landscape of the IPR for $W/t=0.5$ and $W/t=3$, respectively. We emphasize that the blue colored region in Fig.~\ref{fig: supplespectrum}(a) indicates highly delocalized states, including extended states for sufficiently large $\alpha$. On the other hand, we note that the blue colored region in Fig.~\ref{fig: supplespectrum}(f) is at most critical states as discussed in the main text. Figures~\ref{fig: supplespectrum}(b-e) and (g-j) exhibit the energy spectrum for different $\alpha$ values for $W/t=0.5$ and $W/t=3$, respectively. We find that the states for $W<\vert E\vert\le 2t$ are significantly delocalized for $W<2t$ and $\alpha\ge10$, while the states for $\vert E\vert\le W$ remain intermediate values of the IPR. This supports the presence of the mobility edge at $E_\mathrm{ME}=\pm W$. In contrast, the whole spectrum remains these intermediate values of IPR in the cases of $W/t>2$ [Fig.~\ref{fig: supplespectrum}(g-j)] even for sufficiently large $\alpha$ values, which shows the virtual stealthiness. Nevertheless, for all $W$, we figure out that sufficiently large $\alpha$ gives rise to the delocalization into either critical or extended states within the energy window $W\lesssim\vert E\vert\lesssim2t$. These observations are consistent with the discussion in the main text.

\section{Landscape of generalized fractal dimension for different $W$}
\begin{figure}[h]
    \centering
    \includegraphics[width=0.95\textwidth]{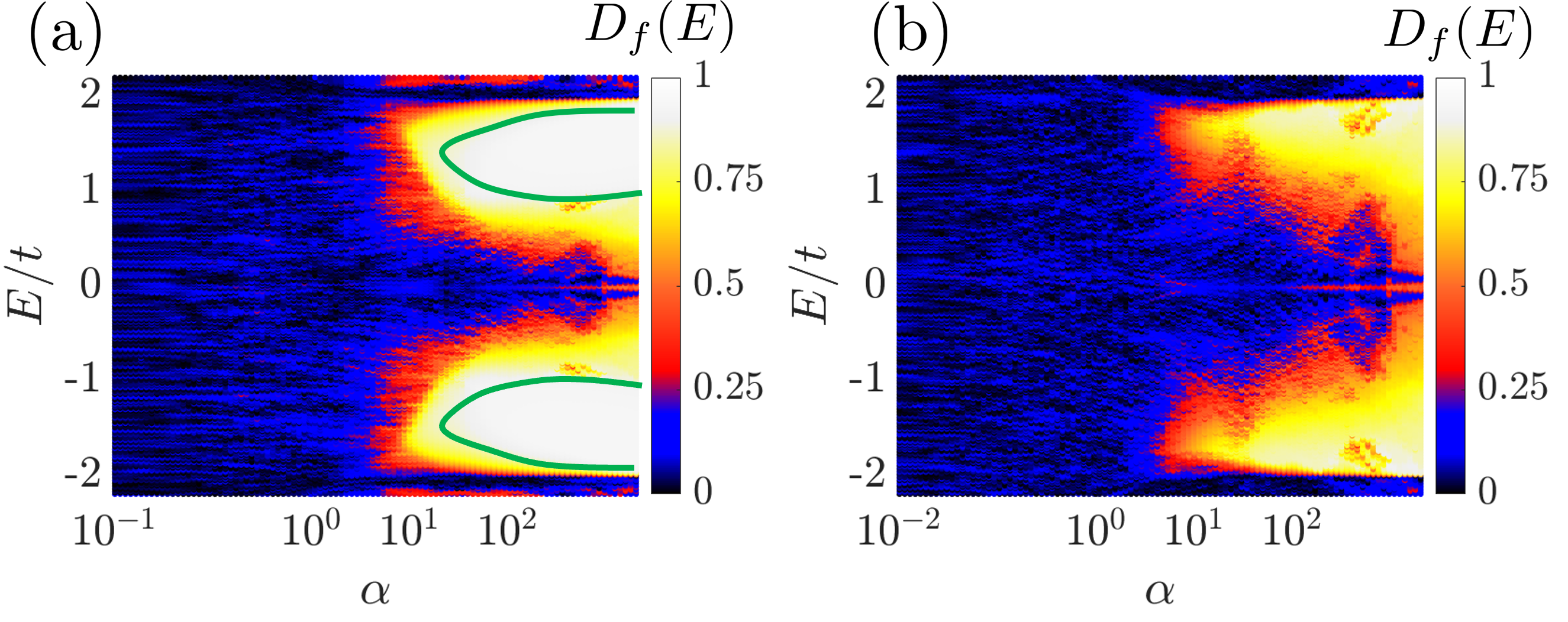}
    \caption{Landscape of generalized fractal dimensions as a function of energy and strength of hyperuniformity, $\alpha$ for $W/t=$ (a) 1 and (b) 2. The green curves in (a) are drawn to emphasize the mobility edge. The system size $2000\le L\le20000$. We use $10^3$ samples of random potentials.}
    \label{fig: supple_fractal}
\end{figure}

Figure~\ref{fig: supple_fractal} exhibits the landscapes of generalized fractal dimension, $D_f(E)$ as a function of $\alpha$, for different $W$ values. For small $\alpha\le1$, $D_f(E)$ is generically small for all energies, that is independent of $W$ [blue colored regions in Fig.~\ref{fig: supple_fractal}]. Since the localization length is independent of the system size in this regime, one can expect that $D_f(E)$ for small $\alpha$ vanishes in the thermodynamic limit.

In contrast, for large $\alpha$, $D_f(E)>0$ is finite in the thermodynamic limit, as they contribute to the (sub)-ballistic transport. In particular, the white region in Fig.~\ref{fig: supple_fractal}(a) indicates the fully extended states, whose $D_f(E)\approx 1$ for $W<\vert E\vert\le2t$. Thus, we have the mobility edges, which approach $E_{\mathrm{ME}}/t=\pm W/t<2$ and $E_{\mathrm{ME}}=\pm2t$ [see green curves in Fig.~\ref{fig: supple_fractal}(a)]. Wheareas, Figure~\ref{fig: supple_fractal}(b) shows that $D_f(E)<1$ for every finite $\alpha$ when $W=2t$. This is consistent with the discussion following Figure 2 in the main text.

\section{Landscape of the scaling behavior of localization length}
    \label{sec:1}
    In this Supplemental Material, we provide the landscape of scaling power of localization length in hyperuniformized disordered potentials as a function of $\alpha$ and energy. Specifically, we investigate $\beta=\frac{d\log (L/\xi)}{d\log L}$, where $\xi$ is the localization length and $L$ is the system size. We use a transfer matrix to compute the localization length. Note that the insulating phase is characterized by $\beta=1$, while the one-dimensional metallic phase with ballistic transport shows $\beta=0$. For intermediate value, $0<\beta<1$, the critical states lead to sub-ballistic transport.

\begin{figure}[h]
    \centering
    \includegraphics[width=0.95\textwidth]{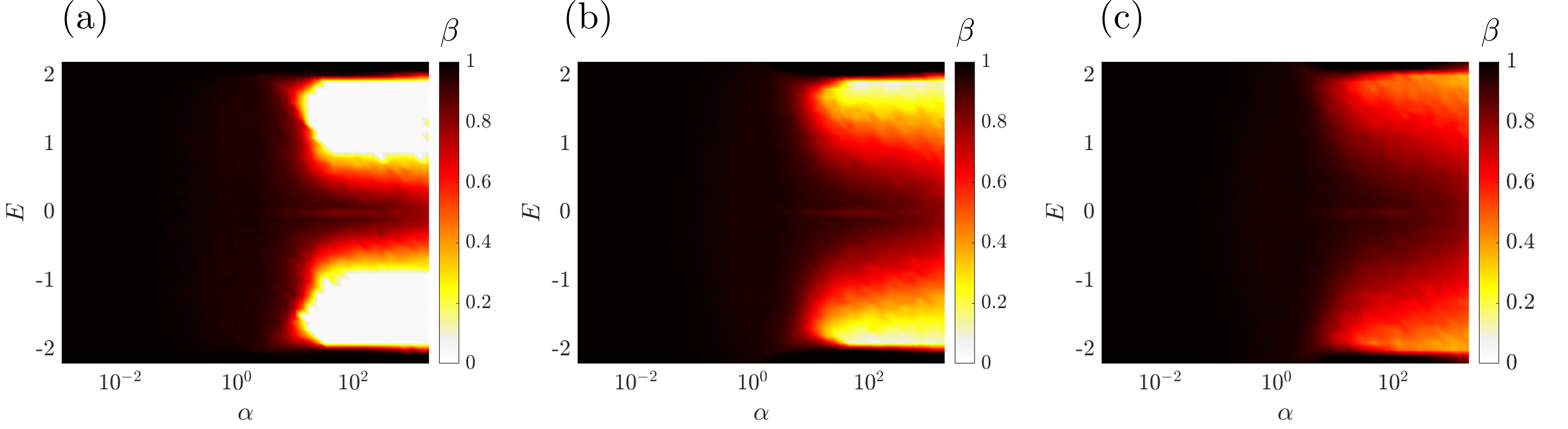}
    \caption{Landscape of scaling power of localization length, $\xi\sim L^{1-\beta}$ as a function of degree of hyperuniformity, $\alpha$, and energy, $E$, for different $W$ values: (a) 1, (b) 2, and (c) 3. Here, $t=1$ and $2000\le L\le 20000$. The number of samples of random potential we used is $10^4$.}
    \label{betalandscape_supple}
\end{figure}

    Figure~\ref{betalandscape_supple} illustrate the landscape of $\beta$ as the function of degree of hyperuniformity, $\alpha$ and energy, $E$. Notably, for $\alpha\le 1$, $\beta=1$ regardless of the energy. This indicates that the class II and III hyperuniformity cannot contribute to the quantum transport. Therefore, in this regime, we expect that the small $D_f$ vanishes in the thermodynamic limit. On the other hand, $\beta<1$ for $\alpha\gg 1$. Particularly, Figure~\ref{betalandscape_supple}(a) shows that $\beta\approx0$ for $W<\vert E\vert <W_c=2t$ for $\alpha\gg 1$ when $W<W_c=2t$. Thus, the system exhibits ballistic transport in this regime despite the presence of random disorder. Although the ballistic transport is absent when $W\ge W_c$, we find sub-ballistic transport, $0<\beta<1$ for around $E\approx 2t$ with sufficiently large $\alpha$, see Figs.~\ref{betalandscape_supple}(b) and (c). Note that these results are consistent with the fractal dimension analysis in the main text.
    
\section{Level spacing statistics analysis}
Let us explore the level spacing statistics across the delocalization transition to clarify the nature of the delocalized states emerging within the energy window $W<\vert E\vert <W_c=2t$ for $\alpha\gg 1$. In detail, we consider the level spacing ratio, $s_n=\frac{\min(\delta_n,\delta_{n+1})}{\max(\delta_n,\delta_{n+1})}$, where $\delta_n=E_{n+1}-E_n$ is the energy spacing. Due to the level repulsion, the average level spacing ratio, $\langle s\rangle$, decreases (increases) as the system size increases when the states are localized (delocalized). In the three-dimensional Anderson model, the localized regime exhibits a Poisson distribution of $s_n$, and hence $\langle s\rangle\approx0.386$. While $\langle s\rangle\approx 0.5307$ for the delocalized regime, as $s_n$ follows the Wigner-Dyson distribution.

\begin{figure}[h]
    \centering
    \includegraphics[width=0.95\textwidth]{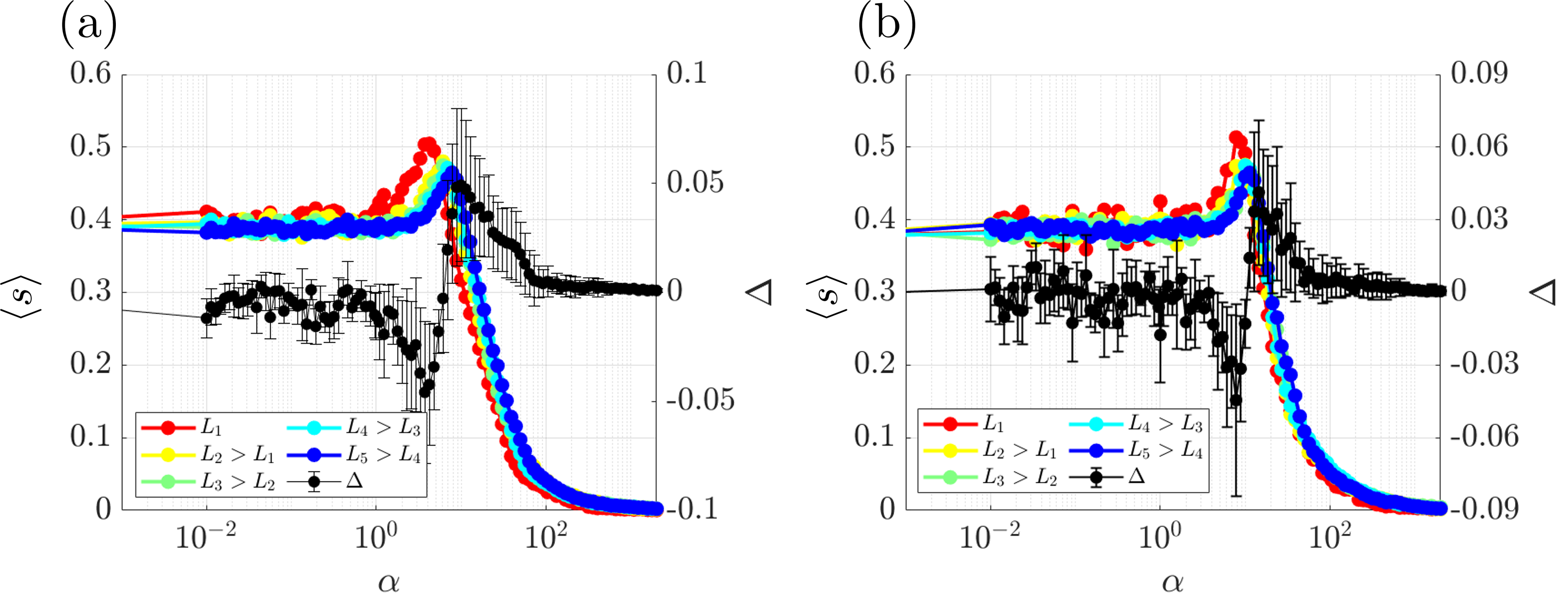}
    \caption{Level spacing statistics of the energy window $W<\vert E\vert <W_c=2t$ as a function of degree of hyperuniformity, $\alpha$ for different system sizes: $L_1=2000$, $L_2=6000$, $L_3=10000$, $L_4=14000$, and $L_5=20000$. The level of disorder of random potentials is (a) $W=0.5t$ and (b) $W=t$, respectively. $t=1$. We use $10^3$ samples of random potentials. The black curves ($\Delta$) are drawn to emphasize the change in $\langle s\rangle$ with system size.}
    \label{fig: levelspacing}
\end{figure}

Figure~\ref{fig: levelspacing} shows $\langle s\rangle$ for different system sizes focusing on the energy window $W<\vert E\vert <W_c=2t$, where the delocalized transition to the ballistic regime characterized by $D_f(E)\approx1$ and $\beta\approx0$ occurs for sufficiently large $\alpha<L$. For small $\alpha$, the level spacing ratio decreases with increasing system size, consistent with localized behavior. While for $\alpha\approx 10$, $\langle s\rangle$ sharply increases with system size, signaling level repulsion. Thus, the change in $\langle s\rangle$ with system size becomes positive (see black curves in Fig.~\ref{fig: levelspacing}). This indicates the delocalization transition. However, unlike the three-dimensional Anderson model—where delocalized states follow Wigner–Dyson statistics—the spacing distribution here collapses into a delta-function form at large $\alpha$ characterized by $\langle s\rangle\approx 0$. This reflects the fact that the delocalized states with fractal dimension $D_f(E)\approx 1$ are momentum-space localized plane waves, rather than ergodic states. %Thus, the transition identified in Fig. 2 marks a crossover from real-space localized states to momentum-space localized plane waves.

\bibliography{reference_supple}
\bibliographystyle{unsrt}